\documentclass[prl,showpacs,twocolumn]{revtex4}

\usepackage{graphics}
\usepackage{amsmath}
\usepackage{slashed}
\usepackage{feynmp}

\begin{document}
\title{Towards computing the lepton and quark mass spectra\\ and their consequences}
\author{Ji\v r\'{\i} Ho\v sek}
\email{hosek@ujf.cas.cz} \affiliation{Department of Theoretical
Physics, Nuclear Physics Institute, Academy of Sciences of the Czech
Republic, 25068 \v Re\v z (Prague), Czech Republic}

\begin{abstract}
We demonstrate that the chiral gauge flavor $SU(3)_f$ dynamics
spontaneously generates the chiral symmetry breaking fermion self
energies $\Sigma(p^2)$ resulting in wide and wild spectra of lepton
and quark masses. The Goldstone theorem then necessarily implies:
(1) Gauge bosons of gauged chiral symmetries absorb the underlying
{\it 'would-be'} Nambu-Goldstone (NG) bosons and become massive.
Masses are determined by $\Sigma(p^2)$s of those fermions to which
they couple, and by the corresponding gauge coupling constants. (2)
Charges of anomalous Abelian chiral fermion currents create
composite mixed-parity {\it pseudo} NG bosons with calculable
couplings to fermions and gauge bosons: (i) the Higgs-looking $h$
with mass due to the $SU(3)_f$ instanton; (ii) the Weinberg-Wilczek
axion $a$ with mass due to the QCD instanton; (iii) the
Anselm-Uraltsev arion $b$ with mass due to the electroweak
instanton. (3) There are the {\it true} composite NG bosons, the
majorons.
\end{abstract}

\pacs{11.15.Ex, 12.15.Ff, 12.60.Fr}

\maketitle

{\it Introduction.} Standard model (SM) Higgs field assisted
generation of lepton and quark masses does not provide much
understanding of their origin. With its 'twenty-some' independently
renormalized and vastly different Yukawa couplings it rather
represents a subtle and successful phenomenology \cite{lee}. Some
even like to argue that the masses of quanta of lepton and quark
fields are the 'environmental' parameters which can acquire any
value.

This is in sharp contrast with understanding the spectra of other
quantum systems. Oscillators, nuclei, atoms and molecules have their
spectra uniquely given by solutions of the Schr\"{o}dinger equation
with scale provided by the corresponding Hamiltonian. In the same
vein, the hadron masses in the paradise (chiral) limit \cite {heiri}
are the calculable multiples of the QCD scale $\Lambda_{QCD} \sim
\rm 200 MeV$. That the laws of QCD at low momenta are known at
present only to computers is another issue.

In this Letter we suggest the asymptotically free dynamics which
arguably provides the calculable spectra of lepton and quark masses.
It is defined by properly gauging the flavor index in SM \cite{ho}:
\begin{eqnarray*}
{\cal L}=-\frac{1}{4}F_{a\mu\nu}F_a^{\mu\nu}+\phantom{bbbbbbbbbbbbb}\\
\bar q_L(3) i\slashed D q_L(3) + \bar u_R(3) i\slashed D u_R(3) +
\bar d_R(\bar 3) i\slashed D d_R(\bar 3)\\ + \bar l_L(\bar 3)
i\slashed D l_L(\bar 3) + \bar e_R(3) i\slashed D e_R(3) + \bar
\nu_R^s(3) i\slashed D \nu_R^s(3)
\end{eqnarray*}
We then demonstrate that by virtue of the Goldstone theorem the
spontaneously generated fermion masses yield the rich spectrum of NG
type excitations with calculable properties. Discussion of other
alternatives to the Higgs mechanism of generating the fermion masses
can be found e.g. in \cite{DEWSB}.

There is one free parameter in ${\cal L}$. It is the dimensionless
gauge coupling constant $h$ hidden in the covariant derivative $D$
and in the gauge field tensor $F$. By dimensional transmutation $h$
turns into theoretically arbitrary scale $\Lambda$ below which the
gauge flavor dynamics is strongly coupled. Because it is not
vector-like like QCD (the chiral fermion fields transform both as
triplets $3$ and antitriplets $\bar 3$ of $SU(3)_f$), we assume
$SU(3)_f$ is not confining. In the following we will discuss how it
behaves in the infrared \cite{appel}. Dealing with the SM chiral
quark and lepton fields the Lagrangian ${\cal L}$ contains also
three $SU(3)_f$ triplets $s=1,2,3$ of the electroweak singlet
(sterile) right-handed neutrino fields $\nu_R^s(3)$. Their existence
is enforced by gauge anomaly freedom \cite{as}. Electroweak
interactions and QCD can be introduced at any moment by gauging in
${\cal L}$ the corresponding fermion indices and adding the gauge
field terms.

{\it Fermion mass generation.} Fermion mass is a bridge between the
left- and right-handed fermion field. Because the octet of flavor
gluons $C_a^{\mu}$ interacts both with the left- and right-handed
lepton and quark fields, its exchanges at low momenta can build up
spontaneously the bridges between the left-handed and right-handed
fermion fields i.e., the fermion masses \cite{njl}.

Fermion masses are given once the chirality-changing fermion proper
self-energies $\Sigma(p^2)$ in the full fermion propagator
\cite{benes} $S(p)^{-1}=\slashed p-\hat \Sigma(p^2)$ are found as UV
finite solutions of the Schwinger-Dyson (SD) homogeneous nonlinear
integral equation \cite{pagels}
\begin{widetext}
\begin{eqnarray}
\hat \Sigma(p)=3\int \frac{d^4k}{(2\pi)^4}\frac{\bar
h^2_{ab}((p-k)^2)}{(p-k)^2}T_a(R)\{\Sigma(k)[k^2+\Sigma^{+}(k)\Sigma(k)]^{-1}P_R+\Sigma^{+}(k)[k^2+\Sigma(k)\Sigma^{+}(k)]^{-1}P_L\}T_b(L)
\label{Sigma}
\end{eqnarray}
\end{widetext}
Here $\hat \Sigma = \Sigma P_L + \Sigma^{+} P_R$, where
$P_{L,R}=\tfrac{1}{2}(1\mp\gamma_5)$ are the chiral projectors. The
fermion mass spectrum is then given by the poles of $S(p)$ i.e., by
solving the equation
\begin{equation}
\rm det[p^2-\Sigma(p^2)\Sigma^{+}(p^2)]=0\label{m}
\end{equation}

For charged leptons, u-quarks and d-quarks $\Sigma(p^2)$ are complex
$3 \times 3$ matrices. The case of neutrinos is more involved and is
discussed separately below. Since $\bar h_{ab}^2(q^2)$ at the
relevant low momenta is not known, finding the fermion spectrum is a
formidable task. In order to proceed we argue as follows:

In the perturbative weak coupling high-momentum region from
$\Lambda$ to $\infty$ which in technical sense guarantees the UV
finiteness of $\Sigma(p^2)$ \cite{pagels} we set $\bar
h_{ab}^2(q^2)$ to zero. {\it The resulting model is thus not
asymptotically, but strictly free above $\Lambda$.} $\Lambda$ of
course should not be confused with a cutoff, simply because there is
nothing we are forced to cut off.

Without loss of generality we fix in the resulting SD equation the
external euclidean momentum as $p=(p,\vec 0)$, integrate over angles
and get
\begin{equation}
\Sigma(p)=\int_0^{\Lambda}k^3dk
K_{ab}(p,k)T_a(R)\Sigma(k)[k^2+\Sigma^{+}\Sigma]^{-1}T_b(L)
\label{Sigmasep}
\end{equation}
where the kernel
\begin{equation}
K_{ab}(p,k)\equiv \frac{3}{4\pi^3}\int_0^{\pi}\frac{\bar
h_{ab}^2(p^2+k^2-2pk \cos \theta)}{p^2+k^2-2pk \cos \theta}\sin^2
\theta d \theta
\end{equation}
is separately symmetric in momenta and in the flavor octet indices.

We suggest to work in separable approximation for the kernel
$K_{ab}(p,k)$, and analyze explicitly the disarmingly simple Ansatz
\begin{equation}
K_{ab}(p,k)=\frac{3}{4\pi^2} \frac{g_{ab}}{pk}\label{sep}
\end{equation}
Separable approximation with a given kernel replaces our ignorance
of knowing the low-momentum $\bar h_{ab}^2(q^2)$ {\it and} of the
low-momentum form of the flavor gluon propagators. Ultimately we
should deal with a system of Schwinger-Dyson equations for several
Green functions. Here $g_{ab}$ is a real symmetric dimensionless
matrix of the effective low-momentum coupling constants. They
reflect the complete breakdown of $SU(3)_f$ and are in principle
calculable. They are analogous to the effective low-momentum
couplings \cite{hl} of chiral perturbation theory of the confining
QCD.

Separable approximation has several further advantages. 1. The
nonlinearity of the integral equation is preserved. We expect that
the nonanalyticity of $\Sigma$ upon the couplings is crucial for
generating the huge fermion mass ratios. 2. In separable
approximation the homogeneous nonlinear integral equation
(\ref{Sigmasep}) is immediately formally solved:
\begin{equation}
\Sigma(p)=\frac{\Lambda^2}{p}T_a(R)\Gamma_{ab}T_b(L)\equiv\frac{\Lambda^2}{p}\sigma
\label{sol}
\end{equation}
where the numerical matrix $\Gamma$ has to fulfil the nonlinear
self-consistency condition (gap equation)
\begin{eqnarray}
\Gamma_{ab}&=&g_{ab}\frac{3}{16\pi^2}\int_{0}^{1}dx (T(R)\Gamma
T(L)) \nonumber\\
&& [x + (T(R)\Gamma T(L))^{+}(T(R)\Gamma T(L))]^{-1}\label{Gamma}
\end{eqnarray}
3. Separable approximation preserves physics contained in $\Sigma$ :
It describes both the fermion masses for a given sort of fermions,
and the fermion mixing, including the CP-violating phases. As in the
Standard model the generally complex $3 \times 3$ matrix $\sigma$
can be put into a positive-definite real diagonal matrix $\gamma$ by
a constant bi-unitary transformation:
\begin{equation}
\sigma=U^{+}\gamma V
\end{equation}
The gap equation becomes
\begin{equation}
\gamma=UT_a(R)U^{+}g_{ab}I(\gamma)VT_b(L)V^{+} \label{gamma}
\end{equation}
where
\begin{eqnarray}
I(\gamma)&=&\frac{3}{16\pi^2}\gamma
\int_{0}^{1}\frac{dx}{x+\gamma^2}= \frac{3}{16\pi^2}\gamma \rm
ln\frac{1+\gamma^2}{\gamma^2}\nonumber\\&\approx&
-\frac{3}{16\pi^2}\gamma \rm ln \gamma^2
\end{eqnarray}
provided $\gamma^2 \ll 1$. The diagonal entries of the equation
(\ref{gamma}) determine the fermion masses, the nondiagonal entries
provide relations for the mixing angles and CP-violating phases.

To account for the possibility of Majorana masses of neutrinos we
introduce the left-handed neutrino field with twelve entries as
\begin{eqnarray}
n_L &=& \left(\begin{array}{c}
\nu_L \\
(\nu_{R})^{{\cal C}}
\end{array}\right)
\end{eqnarray}
and its right-handed partner $(n_L)^{{\cal C}}=n_R$. The general
neutrino chirality-changing proper self-energy $\Sigma_n$ to be
dynamically generated is a  $12 \times 12$ matrix, symmetric by
Pauli principle:
\begin{eqnarray}
\Sigma_n &=& \left(\begin{array}{cc}
\Sigma_{L}     &  \Sigma_{D} \\
\Sigma_{D}^T  &  \Sigma_{R} \\
\end{array}\right)
\label{Sigmanu}
\end{eqnarray}
Here $\Sigma_L$ is a $3 \times 3$ Majorana self-energy of the active
left-handed neutrino fields $\nu_L(\bar 3)$; $\Sigma_R$ is a $9
\times 9$ Majorana self-energy of the sterile left-handed neutrino
fields $(\nu_{R})^{{\cal C}}(\bar 3)$, while $\Sigma_D$ is the
corresponding $3 \times 9$ Dirac neutrino self-energy.

$\Sigma_n$ is also determined by solving the SD equation. In the
neutrino case ${\cal T}_a(n_{R,L})$ are the $12 \times 12$
block-diagonal matrices having on the diagonal four identical $3
\times 3$ blocks of the charged lepton type: ${\cal T}_a(n_L)=\rm
diag(-\tfrac{1}{2}\lambda_a^{*},...)$,${\cal T}_a(n_R)=\rm
diag(\tfrac{1}{2}\lambda_a,...)$.

In separable approximation (\ref{sep}) the neutrino self-energy
$\Sigma_n(p)$ acquires the form
\begin{equation}
\Sigma_n(p) =\frac{\Lambda^2}{p}{\cal T}_a(R)\Gamma_{ab}{\cal
T}_b(L)\equiv\frac{\Lambda^2}{p}\sigma_n \label{soln}
\end{equation}
where the complex numerical  symmetric $12 \times 12$ matrix
$\sigma_n$ can be put into a real diagonal form $\gamma_n$ with
non-negative entries by one unitary $12 \times 12$ matrix $V$:
$\sigma_n = V^{T} \gamma_n V$. The corresponding gap equation which
determines both $\gamma_n$ and $V$ is
\begin{equation}
\gamma_n=V^{*}{\cal T}_a(R)V^{T}g_{ab}I(\gamma_n)V{\cal T}_b(L)V^{+}
\label{gamman}
\end{equation}

{\it Hand-made world of quarks and leptons}. Here we illustrate that
the low-momentum gauge flavor dynamics in the approximation of
separable kernel (\ref{sep}) has the potential of generating the
wide and wild spectrum of quark and lepton masses. Systematic
analysis of fermion mass spectrum and of mixing parameters does not
seem simple, and requires extra work.

1. For the up-type quarks  $T_a(R)=\tfrac{1}{2}\lambda_a$ and
$T_a(L)=\tfrac{1}{2}\lambda_a$. We assume the u-quarks {\it unmixed}
i.e., the matrix $\sigma_u=diag(\gamma_1,\gamma_2,\gamma_3)$, i.e.,
$U_u=V_u=1$ in (\ref{gamma}). We assume further that in the
hand-made world only $g_{33}, g_{38}, g_{88}$ are different from
zero. The diagonal matrix gap equation is easily solved:
$\gamma_i=-\tfrac{1}{16\pi}\alpha_i^{u}\gamma_i \rm ln \gamma_i^2$.
Here $\alpha_1^u=\tfrac{3}{4\pi}(g_{33}+\tfrac{2}{\sqrt
3}g_{38}+\tfrac{1}{3}g_{88})$,
$\alpha_2^u=\tfrac{3}{4\pi}(g_{33}-\tfrac{2}{\sqrt
3}g_{38}+\tfrac{1}{3}g_{88})$ and
$\alpha_3^u=\tfrac{3}{4\pi}\tfrac{4}{3}g_{88}$.

For fermions $i=1, 2, 3$ the fermion mass $m_i$ is defined in terms
of $\Sigma_i(p)=(\Lambda^2/p)\gamma_i$ as $m_i=\Sigma(p^2=m_i^2) =
\Lambda \gamma_i^{1/2}$. Consequently, for the $f=u$-type quark
masses $m_i^u$ we arrive at an appealing mass formula
\begin{equation}
m_i^f=\Lambda \phantom{b} \rm exp (-4\pi/\alpha_i^f)\label{mass}
\end{equation}
So far $\Lambda$ is arbitrary. As we will see later the most
stringent phenomenological constraint on its numerical value comes
from the invisibility of the axion which the model predicts. From
this datum we fix $\Lambda \sim 10^9 \rm GeV$.

2. For the down-type quarks $T_a(R)=-\tfrac{1}{2}\lambda_a^{*}$,
$T_a(L)=\tfrac{1}{2}\lambda_a$. For unmixed down-type quarks with
the coupling matrix $g_{ab}$ already fixed above the formula for
$m_i^d=\Lambda \phantom{b} \rm exp (+4\pi/\alpha_i^d)$ with
$\alpha_i^d=\alpha_i^u$ would describe d-quarks with huge masses. In
general, because $3$ and $3^{*}$ are the inequivalent
representations of $SU(3)_f$, there is no such a unitary matrix $U$
in (\ref{gamma}) which would turn all eight generators
$-\tfrac{1}{2}\lambda_a^{*}$ into $\tfrac{1}{2}\lambda_a$. It is,
however, conceivable that the fermion mixing helps. Indeed, in the
following we show that with simple mixing matrices $U_d$ and $V_d$
we arrive at a sensible mass formula for $m_i^d$.

Consider in the gap equation (\ref{gamma}) for the d-quark masses
the $3 \times 3$ unitary matrices $U,V$
\begin{eqnarray}
U_d = i\left(\begin{array}{ccc}
1     &  0   &   0 \\
0     &  0   &   1 \\
0     &  1   &   0 \\
\end{array}\right) \quad &,& \quad
V_d = i\left(\begin{array}{ccc}
0     &  1   &   0 \\
1     &  0   &   0 \\
0     &  0   &   1 \\
\end{array}\right) \quad
\end{eqnarray}
We arrive at the same exponential mass formula (\ref{mass}) for
$f=d$ with $\alpha_1^d=\tfrac{3}{4\pi}(g_{33}-\tfrac{1}{3}g_{88})$,
$\alpha_2^d=\tfrac{3}{4\pi}(\tfrac{2}{\sqrt
3}g_{38}+\tfrac{2}{3}g_{88})$ and
$\alpha_3^d=\tfrac{3}{4\pi}(-\tfrac{2}{\sqrt 3}g_{38}+\tfrac{2}{
3}g_{88})$.

3. For the charged leptons $(e, \mu, \tau)$
$T_a(R)=\tfrac{1}{2}\lambda_a$, $T_a(L)=-\tfrac{1}{2}\lambda_a^{*}$.
Not surprisingly, if we rotate the L,R generators by the same
unitary matrices as in the case of the d-quarks, we arrive for the
charged leptons at the same mass formulas: $m_i^l=\Lambda
\phantom{b} \rm exp (-4\pi/\alpha_i^l)$ with
$\alpha_i^l=\alpha_i^d$.

4. For the neutrinos the space of ${\cal T}_a(n_{R,L})$ is $12
\times 12$, only one unitary matrix $V$ is at hand and, apparently,
a 'computer-made' approach is necessary. In this exploratory stage
we assume that the suggested mechanism does generate three light
active neutrinos, and at least one heavy neutrino with mass $M_R
\sim 10^9 \rm GeV$.

For having sensible mass formulas we need
$g_{33}>\tfrac{1}{3}g_{88}>\tfrac{1}{\sqrt 3}g_{38}>0$. The
numerical values of the low-momentum effective couplings come out
reasonable. With $\Lambda=10^9 \rm GeV$ and a quark with mass
$m^q=10^2 \rm GeV$ the corresponding $\alpha^q=4\pi/7 \rm ln 10$
whereas for the charged lepton with mass lighter by five orders of
magnitude, $m^l=1 \rm MeV$ $\alpha^l=4\pi/12 \rm ln 10$. Using the
same formula also for neutrinos, another six orders of magnitude for
$m^{\nu}=1 \rm eV$ turns into $\alpha^{\nu}=4\pi/18 \rm ln 10$. It
is natural to associate the ratios $\alpha^l/\alpha^q=7/12$ and
$\alpha^{\nu}/\alpha^l=2/3$ with mixing angles entering the general
formulas (\ref{gamma}) and (\ref{gamman}).

{\it Fate of gauged chiral symmetries.} These are the $SU(2)_L
\times U(1)_Y$ in which the gauge bosons $W, Z, A$ couple to
electroweakly interacting fermions, and the $SU(3)_f$ gauge flavor
dynamics in which the flavor gluons $C$ couple to electroweakly
interacting fermions {\it and} to sterile right-handed neutrinos.
Dynamically generated fermion self-energies $\Sigma(p^2)$ break
spontaneously the symmetry $SU(2)_L \times U(1)_Y$ down to unbroken
$U(1)_{em}$, and the symmetry $SU(3)_f$ completely. As a consequence
the composite {\it non-Abelian} 'would-be' NG bosons become the
longitudinal polarization states of $W,Z$ and of all eight $C$
bosons. General arguments of the renown BEH mechanism \cite{beh}
apply: First, the composite 'would-be' NG bosons are explicitly
visualized as massless poles in Ward-Takahashi (WT) identities for
the corresponding vertex functions \cite{jj} with residues
proportional to $\Sigma(p^2)s$. In reality, there are always several
fermions which incoherently contribute, and there is a mixing
\cite{hos}. For simplicity it is neglected here. In the electroweak
case neglecting all fermion masses against the top quark mass
$m_{\rm top}$ is a very good approximation. In the case of $SU(3)_f$
we consider only the heaviest right-handed Majorana mass $M_R$.
Second, the gauge boson mass squared, defined according to Schwinger
\cite{sch} as a residue at the massless pole of the gauge boson
polarization tensor is computed in terms of $\Sigma(p^2)$ by the
Pagels-Stokar formula \cite{ps} used in the present context in
technicolor \cite{tc}
\begin{equation}
F^2=8N\int\frac{d^4p}{(2\pi)^4}\frac{\Sigma^2(p^2)-\tfrac{1}{4}p^2(\Sigma^2(p^2))^{'}}{(p^2+\Sigma^2)^2}
\label{ps}
\end{equation}
where $N$ is a loop factor. Having the explicit form of
$\Sigma_f(p^2)=(\Lambda^2/p)\gamma_f\equiv m_f^2/p$ we have also the
explicit form of $F^2=\tfrac{5}{16\pi} N m_f^2$.

Neglecting the fermion mixing we conclude that \cite{hos}
\begin{eqnarray*}
m_W^2 & \simeq &  \tfrac{1}{4}g^2\tfrac{5}{16\pi}N m_{\rm top}^2\\
 m_Z^2 & \simeq & \tfrac{1}{4}(g^2 + g'^{2})
\tfrac{5}{16\pi}N m_{\rm top}^2.
\end{eqnarray*}
For $N=n_c \times n_f \times 2=3 \times 3 \times 2 =18$ the
numerical agreement of $F \sim 231.5 \rm GeV$ with the SM value
$v=246 \rm GeV$ is suspiciously good. Canonical SM tree-level
relation $m_W^2/m_Z^2 \rm cos^2\theta_W=1$ would be exact for
$m_{\rm bottom}= m_{\rm top}$ \cite{hos}.

Analogously, for strongly coupled flavor gluons with effective
couplings to fermions of order one we estimate the masses of color
gluons: $m_C^2\simeq  M_R^2$. Because $M_R \sim \Lambda=10^9 \rm
GeV$ the flavor-changing flavor gluons $C$ have no observable
effects in SM FCNC processes characterized by a typical effective
strength $G_F^2 m_{\rm top}^2$.

{\it Fate of global chiral symmetries.} First we observe that the
Lagrangian ${\cal L}$ with three triplets ($s=1,2,3$) of sterile
$\nu_R^s$ obeys the global $U(3)_s=U(1)_s \times SU(3)_s$ {\it
sterility} symmetry. If all sterile neutrinos acquire dynamically
the Majorana masses there must be an octet of composite massless NG
majorons \cite{peccei}. Having by virtue of the WT identity the
direct (derivative) couplings only to sterile neutrinos for particle
physics the majorons should be harmless.

It remains to discuss six Abelian symmetries generated by charges of
six chiral fermion currents $j^{\mu}_i, i=q_L, u_R, d_R, l_L, e_R,
\nu_R$, taking into account {\it the quantum effects of axial
anomalies}. Dealing with four gauge forces we have \cite{thooft}
\begin{eqnarray*}
\partial_{\mu}j_{q_L}^{\mu} & = & \partial_{\mu}(\bar q_L \gamma^{\mu} q_L)  =  -A_Y-9A_W-6A_G-6A_F\\
\partial_{\mu}j_{u_R}^{\mu} & = & \partial_{\mu}(\bar u_R \gamma^{\mu} u_R)  =  8A_Y+3A_G+3A_F \\
\partial_{\mu}j_{d_R}^{\mu} & = & \partial_{\mu}(\bar d_R \gamma^{\mu} d_R)  =  2A_Y+3A_G+3A_F \\
\partial_{\mu}j_{l_L}^{\mu} & = & \partial_{\mu}(\bar l_L \gamma^{\mu} l_L)  =  -3A_Y-3A_W-2A_F \\
\partial_{\mu}j_{e_R}^{\mu} & = & \partial_{\mu}(\bar e_R \gamma^{\mu} e_R)  =  6A_Y+A_F \\
\partial_{\mu}j_{\nu_R}^{\mu} & = & \partial_{\mu}(\bar\nu_R\gamma^{\mu}\nu_R)  =  3A_F
\end{eqnarray*}
where $A_X=\tfrac{g_X{2}}{32\pi^2}F_X\tilde F_X$, and $X$
abbreviates the gauge forces $U(1)_Y$, $SU(2)_L$, $SU(3)_c$ and
$SU(3)_f$, respectively.

There are two anomaly-free currents parameterized by two real
parameters $e,f$: $j^{\mu}_{e,f} =
-\tfrac{1}{6}(e+3f)j_{q_L}^{\mu}+\tfrac{1}{3}(-2e+3f)j_{u_R}^{\mu}+\tfrac{1}{3}(e-6f)j_{d_R}^{\mu}+
\tfrac{1}{2}(e+3f)j_{l_L}^{\mu}+ej_{e_R}^{\mu}+fj_{\nu_R}^{\mu}$.

For $f=0$, $e=-2$ we get the anomaly free current of weak
hypercharge $Y$, $j^{\mu}_Y$. Because the electric charge $Q$ is
$Q=I_3 + \tfrac{1}{2}Y$, the hypercharge current is hidden in the
vectorial electromagnetic current coupled to the massless photon
field $A$, and does not create any 'would-be' NG boson.

For definiteness we fix the other anomaly free current
$j^{\mu}_{Y'}$ by $f\neq0, e=0$:
\begin{eqnarray*}
\tfrac{1}{f}j^{\mu}_{Y'} =
-\tfrac{1}{2}j_{q_L}^{\mu}+j_{u_R}^{\mu}-2j_{d_R}^{\mu}+
\tfrac{3}{2}j_{l_L}^{\mu}+j_{\nu_R}^{\mu}
\end{eqnarray*}
It creates the true fermion-antifermion massless composite NG boson.
Because of its $\nu_R$ component its couplings with fermions are
tiny. To prevent any conflict with data we better gauge the
corresponding Abelian symmetry. The NG boson becomes 'would-be' and
the new $Z'$ heavy, with mass mass $m_{Z'} \sim g'' M_R$ where $g''$
is a new gauge coupling constant.

One of four anomalous currents is the {\it vectorial} baryon current
$j^{\mu}_B=\tfrac{1}{3}(j^{\mu}_{q_L}+j^{\mu}_{u_R}+j^{\mu}_{d_R})$
which does not create any pseudo NG boson.

{\it Pseudo NG bosons.} We are left with three anomalous Abelian
chiral currents which create three pseudo NG bosons with masses due
non-Abelian anomalies by instantons \cite{inst}. We identify them
with the Higgs-looking resonance $h$, the Weinberg-Wilczek axion $a$
\cite{ww}, and the Anselm-Uraltsev arion $b$ \cite{au},
respectively. The Abelian anomaly $A_Y$ is expected not to play any
role in giving masses to the pseudo NG bosons. The Abelian gauge
field $B^{\mu}$ is topologically trivial, and there is no instanton
associated with it.

There is a freedom in fixing the coefficients in $j_h^{\mu}=\sum_i
h_i j^{\mu}_i$, $j_a^{\mu}=\sum_i a_i j^{\mu}_i$, and
$j_b^{\mu}=\sum_i b_i j^{\mu}_i$, $i=q_L, u_R, d_R, l_L, e_R,
\nu_R$. It can be employed for properly fixing the data. For
definiteness we demand: 1. The charges of four currents
$j^{\mu}_{Y'}, j_h^{\mu}, j_a^{\mu}, j_b^{\mu}$ creating the
'would-be' and pseudo NG type excitations should be orthogonal. 2.
Higgs does not interact with gluons, hence, $\partial_{\mu}
j_h^{\mu}$ does not contain $A_G$. 3. The requirement $h_{\nu_R}=0$
implies that the interactions of $h$ with fermions are not
suppressed by $M_R$. 4. The axion $a$ is invisible, hence
$a_{\nu_R}\neq 0$. 5. The axion interacts only strongly. Hence,
$\partial_{\mu} j_a^{\mu}$ does not contain $A_F, A_W, A_Y$. 6. The
arion interacts only electroweakly. Hence, $\partial_{\mu}
j_b^{\mu}$ does not contain $A_F$ and $A_G$. 7. To get the currents
completely fixed we set, without any obviously good reason,
$a_{e_R}=b_{q_L}=0$.

The result is
\begin{eqnarray*}
\tfrac{1}{h}j_h^{\mu}& = &
j_{q_L}^{\mu}+\tfrac{8}{5}j_{u_R}^{\mu}+\tfrac{2}{5}j_{d_R}^{\mu}-
\tfrac{1}{5}j_{l_L}^{\mu}-\tfrac{62}{145}j_{e_R}^{\mu}\\
\tfrac{1}{a}j_a^{\mu}& = &
j_{q_L}^{\mu}-\tfrac{7}{12}j_{u_R}^{\mu}-\tfrac{5}{3}j_{d_R}^{\mu}-3j_{l_L}^{\mu}+\tfrac{9}{4}j_{\nu_R}^{\mu}\\
\tfrac{1}{b}j_b^{\mu}& = &j_{u_R}^{\mu}-j_{d_R}^{\mu}-\tfrac
{8}{9}j_{l_L}^{\mu}+\tfrac{29}{9}j_{e_R}^{\mu}-\tfrac{5}{3}j_{\nu_R}^{\mu}\\
\end{eqnarray*}
The normalization factors $f, h, a, b$ are the arbitrary real
numbers.

{\it Unique form of the pseudo NG currents fixes the effective
interactions of the pseudo NG bosons.}  1. The WT identities
associated with the pseudo NG currents determine the effective
Yukawa couplings of the pseudo NG bosons with fermions. Clearly,
although proportional to the fermion masses they are not universal
as in the Standard model. Moreover, in separable approximation they
are pseudoscalar. 2. The divergences of the pseudo NG currents fix
the effective interactions of the pseudo NG bosons $h, a, b$ with
the respective gauge fields. There is no way how $h$ could interact
with $WW$ and $ZZ$ by the renormalizable SM interactions.

{\it The effective interactions of the pseudo NG bosons with
fermions and non-Abelian gauge bosons give rise to their masses}
\cite{ww}. For the axion $a$ massive due to the QCD instanton we use
the formula \cite{ww} $m_a \sim \Lambda_{QCD}^2/\Lambda \sim 10^{-2}
\rm eV$ from which we fix the scale $\Lambda \sim 10^9 \rm GeV$. For
the electroweak axion $b$ (arion) massive due to the screened
electroweak instanton we use the formula \cite{au2}
\begin{equation}
m^2_b=Cm_W^2(8\pi^2/g_W^2)^4 \rm exp(-8\pi^2/g_W^2) \label{au}
\end{equation}
with $C<1$. It results in $m_b=3.10^{-28}\rm eV C^{1/2}$. For the
$m_h$ we use the formula (\ref{au}) with $m_W$ replaced by $m_C \sim
10^9 \rm GeV$, and $g_W$ replaced by a bigger $h$. A wishful
estimate is then $m_h \sim 10^2 \rm GeV$.

{\it Conclusion.} There are two main results of this Letter. First,
we have suggested the explicit framework for computing the mass
spectra of leptons and quarks and their mixing parameters within the
ultraviolet-complete chiral non-Abelian $SU(3)_f$ gauge flavor
dynamics. Systematic analysis remains to be done. Second, once the
the chirality-changing fermion proper self-energies $\Sigma(p^2)$
are spontaneously generated, the Goldstone theorem provides definite
predictions. Their detailed analysis also requires further work.

1. There must be a composite pseudo NG boson $h$ with mass due to
the anomaly of gauge flavor dynamics. We boldly suggest to identify
it with the $126 \rm \phantom a GeV$ resonance recently discovered
by the LHC at CERN \cite{higgs}. Such a particle is not, however,
the standard SM Higgs. First, our $h$ is composite. The
compositeness scale, though, is very large. Second, the mass $m_h$
has nothing to do with the Fermi scale. Third, in accord with a
theorem \cite{vW} $h$ is definitely not a pure CP-even scalar like
the SM Higgs. Fourth, the Yukawa couplings of $h$ with fermions are
in a specific way proportional to the fermion masses. Fifth, the
couplings of $h$ to $WW$ and $ZZ$ are not of the SM type. The
dimension-5 couplings are fixed.

2. There must be a light composite pseudo NG boson which we identify
with the Weinberg-Wilczek QCD axion. Its invisibility fixes the
scale of $SU(3)_f$, $\Lambda \sim 10^9 \rm GeV$.

3. There must be an ultra-light composite electroweak axion $b$, and
an octet of massless composite majorons $m_a$. The very light or
massless, very weakly interacting particles, collectively
abbreviated as WISPs (weakly interacting slim particles) are subject
to intense experimental searches \cite{ringwald}.

4. The gauge bosons of gauged chiral symmetries acquire masses
proportional to the sum of masses of fermions to which they couple.
This in particular implies that there is no genuine electroweak
Fermi scale. The masses $m_W$ and $m_Z$ are the unique consequence
of a sum rule essentially saturated by the top quark mass computed
in gauge flavor dynamics.

For theoretical consistency there must be, besides three SM active
$\nu_L$, nine sterile $\nu_R$. While computing the mass spectra of
quarks and charged leptons is at present "merely" a post-diction
(yet to be made), predicting the spectrum of neutrino masses is a
crystalline challenge.

{\it Outlook.} The new dynamics, becoming strong at a high energy
scale $\Lambda$ should, besides generating masses of its elementary
excitations i.e. leptons, quarks and flavor gluons, generate also
entirely new collective excitations. Those guaranteed by the
existence theorem of Goldstone are discussed above. The very heavy
composites might also be useful. Why, first of all, the origin of
dark matter should not be the same as of the shining one ? The
analogs of baryons, fermionic {\it sterile} neutrino composites
$\epsilon^{abc}\nu_{aR}\nu_{bR}\nu_{cR}$ seem to be natural
candidates for the 'co-generated' dark matter \cite{barr}.
\begin{acknowledgments}
I am grateful to Petr Bene\v s, Tom\'a\v s Brauner and Adam Smetana
for valuable discussions and joyful collaboration.
\end{acknowledgments}

\end{document}